\documentclass{aa}
\usepackage{graphicx}
\usepackage{txfonts}
\begin{document}

   \title{Shedding light on the formation mechanism of the shell galaxy NGC474 with MUSE\thanks{Based on data from ESO program 099.B-0328(A) (PI: Fensch).}}

   \author{J\'er\'emy Fensch
          \inst{1,2}  
          \and
          Pierre-Alain Duc\inst{3}
          \and
          Sungsoon Lim\inst{4}
          \and 
          \'Eric Emsellem\inst{2,5}
          \and
          Michal B\'ilek\inst{3}
          \and
          Patrick Durrell\inst{6}
          \and
          Chengze Liu\inst{7}
          \and
          Éric Peng\inst{8,9}
          \and
          Rory Smith\inst{10}
          }

   \institute{
   Univ. Lyon, ENS de Lyon, Univ. Lyon 1, CNRS, Centre de Recherche Astrophysique de Lyon, UMR5574, F-69007 Lyon, France
   email: jeremy.fensch@ens-lyon.fr	
   \and 
   European Southern Observatory, Karl-Schwarzschild-Str. 2, D-85748 Garching, Germany 
   \and 
   Observatoire Astronomique de Strasbourg (ObAS), Universit\'e de Strasbourg - CNRS, UMR 7550 Strasbourg, France
   \and
   University of Tampa, 401 West Kennedy Boulevard, Tampa, FL 33606, USA
   \and
   Univ. Lyon, Univ. Lyon1, ENS de Lyon, CNRS, Centre de Recherche Astrophysique de Lyon UMR5574, F-69230, Saint-Genis-Laval, France
   \and
   Department of Physics \& Astronomy, Youngstown State University, Youngstown, OH 44555     
   \and
   Department of Astronomy, School of Physics and Astronomy, and Shanghai Key Laboratory for Particle Physics and Cosmology, Shanghai Jiao Tong University, Shanghai 200240, China
   \and
Department of Astronomy, Peking University, Yi He Yuan Lu 5, Hai Dian District, Beijing 100871, People's Republic of China
\and
	Kavli Institute for Astronomy \& Astrophysics and Department of Astronomy, Peking University, Yi He Yuan Lu 5, Hai Dian District, Beijing 100871, People's Republic of China
\and
Korea Astronomy and Space Science Institute (KASI), 776 Daedeokdae-ro, Yuseong-gu, Daejeon 34055, Korea
   }

   \date{}

 
  \abstract
  {
   Stellar shells around galaxies could provide precious insights into their assembly history. However, their formation mechanism remains poorly empirically constrained, in particular the type of galaxy collisions at their origin. We present MUSE@VLT data of the most prominent outer shell of NGC~474, to constrain its formation history.
   The stellar shell spectrum is clearly detected, with a signal-to-noise ratio of around 65 pix$^{-1}$. We use a full spectral fitting method to determine the line-of-sight velocity and the age and metallicity of the shell and associated point-like sources within the MUSE field of view.
  We detect six globular cluster (GC) candidates and eight planetary nebula (PN) candidates which are all kinematically associated to the stellar shell. We show that the shell has an intermediate metallicity, [M/H] = $-0.83^{+0.12}_{-0.12}$ and a possible $\alpha$-enrichment, [$\alpha$/Fe] $\sim$ 0.3. Assuming the material of the shell comes from a lower mass companion, and that the latter had no initial metallicity gradient, such a stellar metallicity would constrain the mass of the progenitor to be around $7.4\times10^8$~M$_\odot$, implying a merger mass ratio of about 1:100. However our census of PNs and earlier photometry of the shell would suggest a much higher ratio, around 1:20. Given the uncertainties, this difference is significant only at the $\simeq 1\sigma$ level. We discuss the characteristics of the progenitor, in particular whether the progenitor could also be composed of stars from the low metallicity outskirts from a more massive galaxy. Ultimately, the presented data does not allow us to put a firm constraint on the progenitor mass. 
We show that at least two globular cluster candidates possibly associated with the shell are quite young, with ages below 1.5~Gyr. We also note the presence of a young ($\sim$ 1~Gyr) stellar population in the center of NGC 474. The two may have resulted from the same event. 
}


   \maketitle
%
\section{Introduction}

According to the current cosmological paradigm, galaxies assemble through a continuous process of accretion of gas and successive merging with other galaxies \citep{White78}. This merging history can leave low-surface brightness (LSB) imprints in the halo of galaxies, in the shape of stellar streams, plumes, tidal tails or shells \citep[see e.g.][]{Mihos05, Martinez10, vDokkum14, Duc15, Mancillas19simu, Mueller19}. Thus, the study of these features can help reconstructing the assembly history of galaxies \citep[see e.g.][]{Foster14, Longobardi15}, as well as even serve as probes for gravity in the low acceleration regime \citep[see e.g.][]{Ebrova12, Bilek13}. \\

Young prominent tidal tails are mostly gas-rich and can be analyzed through their gas component \citep[see e.g.][]{Yun94, Duc00, Williams02}. However, plumes, streams and shells are relatively gas poor \citep[but see][]{Charmandaris00} and only their faint stellar absorption lines are usually available for spectroscopy. An alternative to study the dynamics and chemical composition of gas-poor tidal features is to use bright point sources, such as globular clusters (GC) or planetary nebulae (PN) to study the dynamics and chemical composition of plumes or streams \cite[see e.g.][]{Durrell03, Mullan11, Forbes12, Blom14, Foster14}. \\

   \begin{figure*}[ht]
   \centering
   \includegraphics[angle=0,width=18cm]{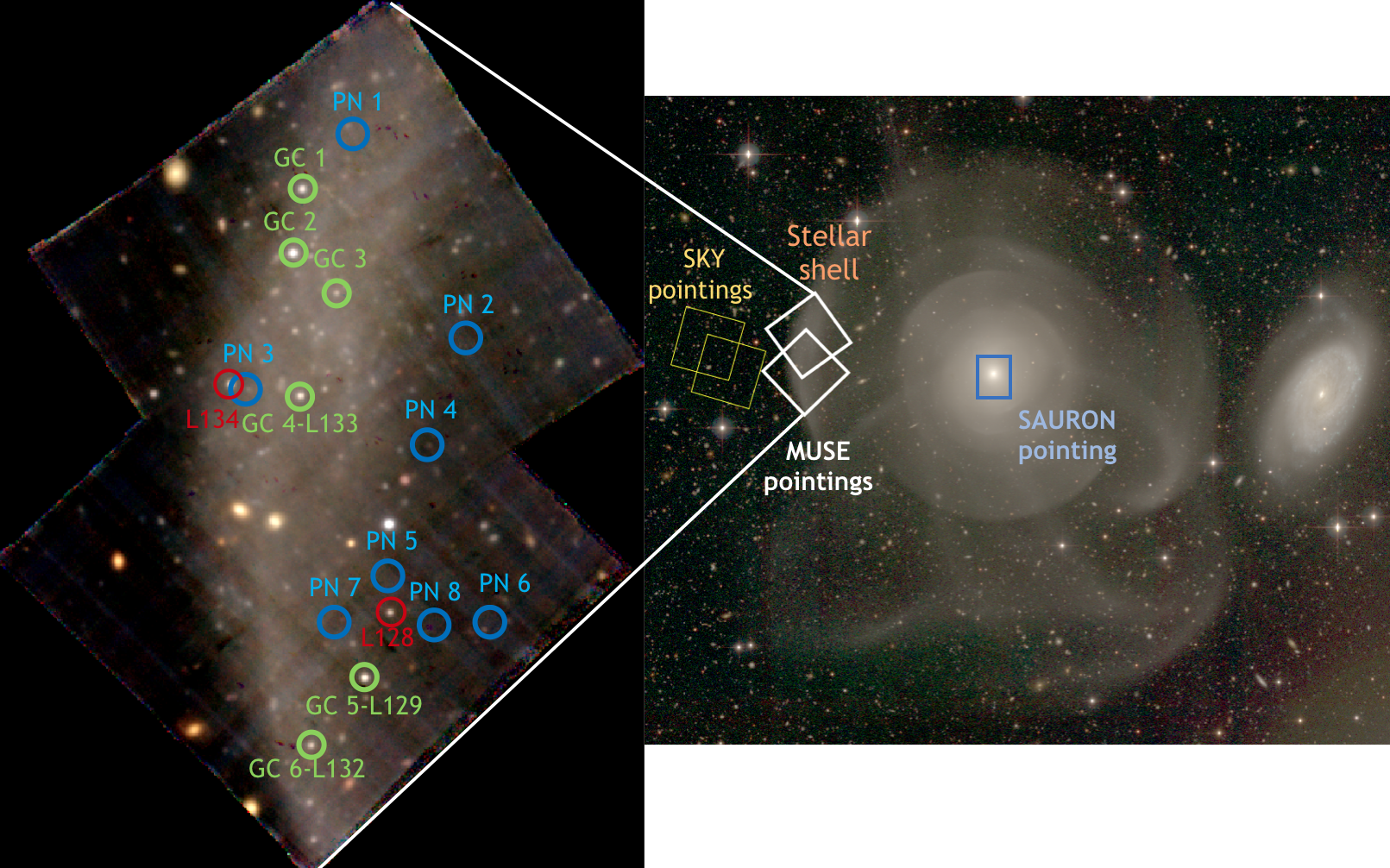}
      \caption{{\it Left}: True color image constructed from the MUSE data using as colors the g,r and i filters of Megacam@CFHT. The locations of the detected PNe are shown with blue circles. In green are shown the detected GCs. GC~4, 5 and 6 were part of L17 sample, their number in that catalog is given after their name. The two red circles show detections that were classified by L17 as GC candidates but are not confirmed with the present data. {\it Right}: Zoom-out true color image from CFHT (g,r,i bands from \citealt{Duc15}). The MUSE and SKY pointings are shown  respectively in white and yellow. The SAURON pointing from \citet{deZeeuw02} is shown in blue. The MUSE and SAURON pointings are respectively 1'x1' and 33\arcsec x 41\arcsec wide. Assuming a distance to NGC~474 of 30.9~Mpc \citep{Cappellari11}, 1' encompasses around 8.9~kpc. North is up and East is right.
              }
         \label{fig:map}
   \end{figure*}

Stellar shells are relatively frequent around elliptical and lenticular galaxies ($\sim$ 20\%, \citealt{Tal09, Duc15, Pop18}, but see \citealt{Krajnovic11}) and have relatively high surface brightness for LSB structures (up to 25~mag~arcsec$^{-2}$, \citealt{Johnston08, Atkinson13}) compared to streams and plumes. They thus appear as convenient tracers of past merging events. Shells systems have been classified within three types, corresponding to the distribution of the shells position angles: Type I for aligned shells, Type II for random orientation and Type III for ambiguous or undetermined orientations \citep{Wilkinson87}. \\

From numerical simulations, it could be inferred that the mechanisms responsible for shell formation are likely to be near-radial infalls of relatively massive galaxies \citep[mass ratio above 1:10, see e.g.][]{Quinn84, Hernquist92, Pop18, Karademir19}, while mergers involving higher angular momentum tend to create stellar streams \citep{Amorisco15, David15}. It should be noted that high mass ratio are favored as higher mass progenitors should be able to create shells from a wider range of impact parameters \citep{Pop18}. Shells are therefore thought to be made up of stars from the accreted and tidally disrupted satellite, at the apocenter of their orbit around the host galaxy. However, the spectrophotometric data needed to test these scenarios are scarce, mostly because they are extremely challenging to obtain \citep[but see longslit spectra from][]{Pence86}. Moreover, this paucity makes that the extent to which point sources - namely GCs and PNe - could trace the kinematics of these stellar structures is still unknown. \\

In this study, we propose to empirically constrain the formation scenario of stellar shells by using the revolutionary capabilities of MUSE on one of the most spectacular shell-rich systems, NGC~474. This galaxy, classified as peculiar by \citet{Arp66} under the name Arp~227, is either a lenticular galaxy or a fast rotating elliptical \citep{Hau96,Emsellem04}, located at 30.9~Mpc \citep{Cappellari11}. The deepest imaging data obtained to date on this system revealed a total of at least ten concentric shells and six radial streams up to 5.6$\arcmin$ from the galaxy \citep[that is 50~kpc, ][ MATLAS survey]{Duc15}. The non-alignment of the position angles of the inner shells around the galaxy (within 100$\arcsec$, that is 15~kpc from the center) classified NGC~474 as a type II shell galaxy \citep{Prieur90, Turnbull99}. These inner shell systems have already been studied via their photometry \citep{Sikkema06, Sikkema07} which revealed the presence of a blue inner shell, a hint for a recent minor merger event. In this paper we present the data obtained on the brightest shell (24.8 mag.arcsec$^{-2}$ in the g-band), located $\sim$30~kpc from the host, which has a few spatially associated GC candidates, selected by color on ultra-deep CFHT imaging \citep[][L17 in the following]{Lim17}. \\

The paper is organized as follows. The data and methods are presented in Section~\ref{sec:data} and the results in Section~\ref{sec:res}. The discussion and conclusion are presented in Section~\ref{sec:disc} and \ref{sec:conc}. All magnitudes are AB magnitudes.

\section{Data and methods}
\label{sec:data}

\subsection{Observations and Reduction steps}

VLT/MUSE observations were conducted in service mode during gray time from November 2017 to November 2018, with \texttt{CLEAR} conditions and observed seeing between 0.5" and 0.9". Observations were split into 16 observing blocks (OBs) of one hour each, amounting to a total 5.1h on-target integration time for each of the two pointings. Each OB was split into four individual on-target exposures with small dithers and 90 degree-rotations to avoid cosmic ray contamination and slicer patterns. All OBs had a "OSOOSO" sequence, where S stands for sky, with each 150~s exposure, and O stands for object, 583~s exposure. Fig.~\ref{fig:map} shows the location of the field of view. We overlapped the two pointing footprints to increase the signal-to-noise on the shell. 

The OBs were all reduced using the latest MUSE \texttt{ESOREX} pipeline recipes \citep[version 2.4.2][]{Weilbacher20}. The reduction follows the standard steps, including sky subtraction. To improve sky subtraction, we use the Zurich Atmosphere Purge software \citep[ZAP, ][]{Soto16}, using the sky available on the left of the shell.  We used 45 eigenvectors and 300 pixels for the window for the continuum subtraction. 

{On Fig.~\ref{fig:map} is also shown the field of view of archival\footnote{available here: http://www-astro.physics.ox.ac.uk/atlas3d/. See SAURON data from Paper I: \citealt{Cappellari11}.} SAURON data centered on NGC~474 \citep{deZeeuw02, Cappellari11}. This data is commented in Section~\ref{disc:trigger}.

\subsection{Detection and spectral extraction}

   \begin{figure*}[ht]
   \centering
   \includegraphics[angle=0,width=18cm]{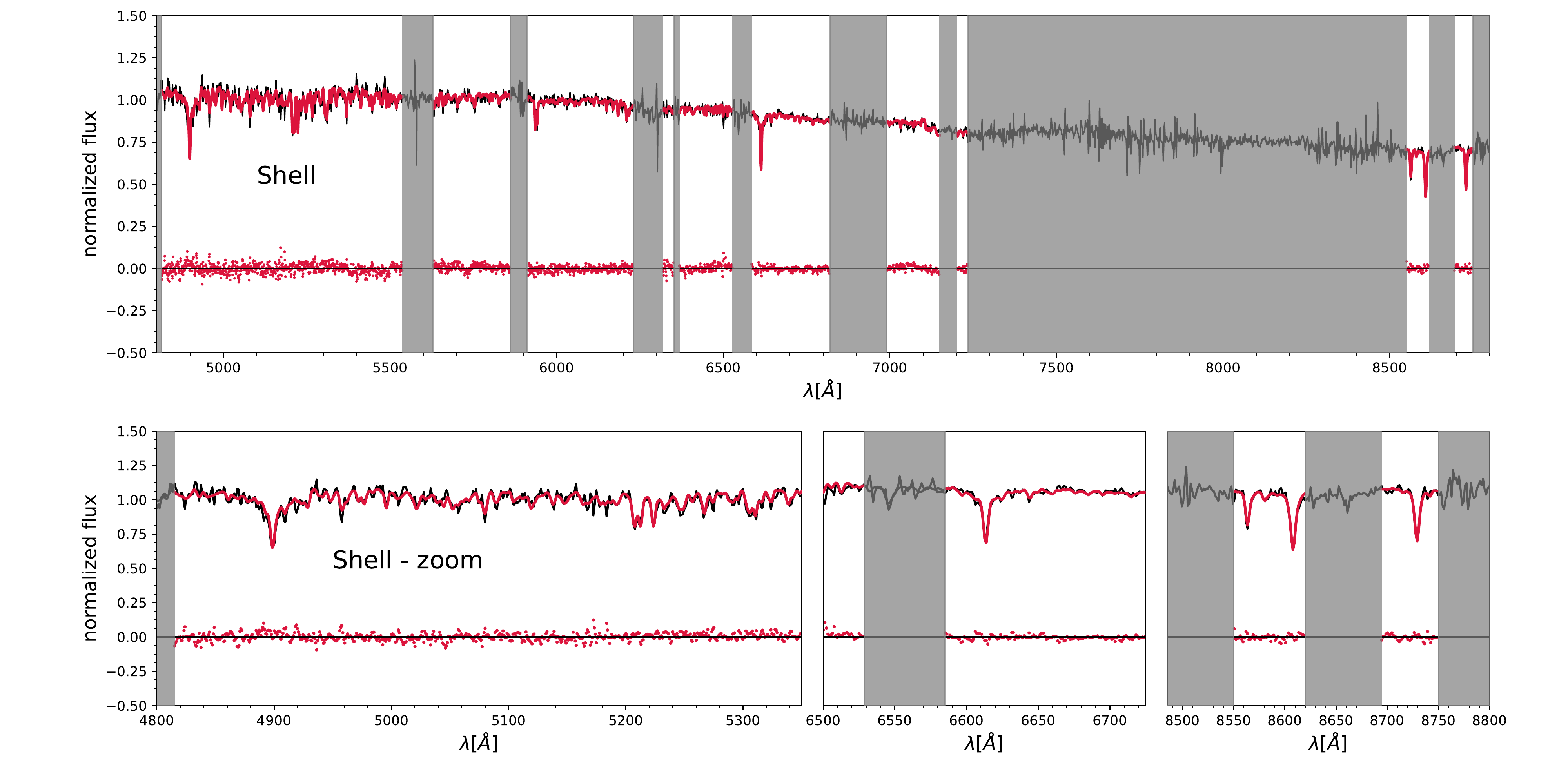}
   \includegraphics[angle=0,width=18cm]{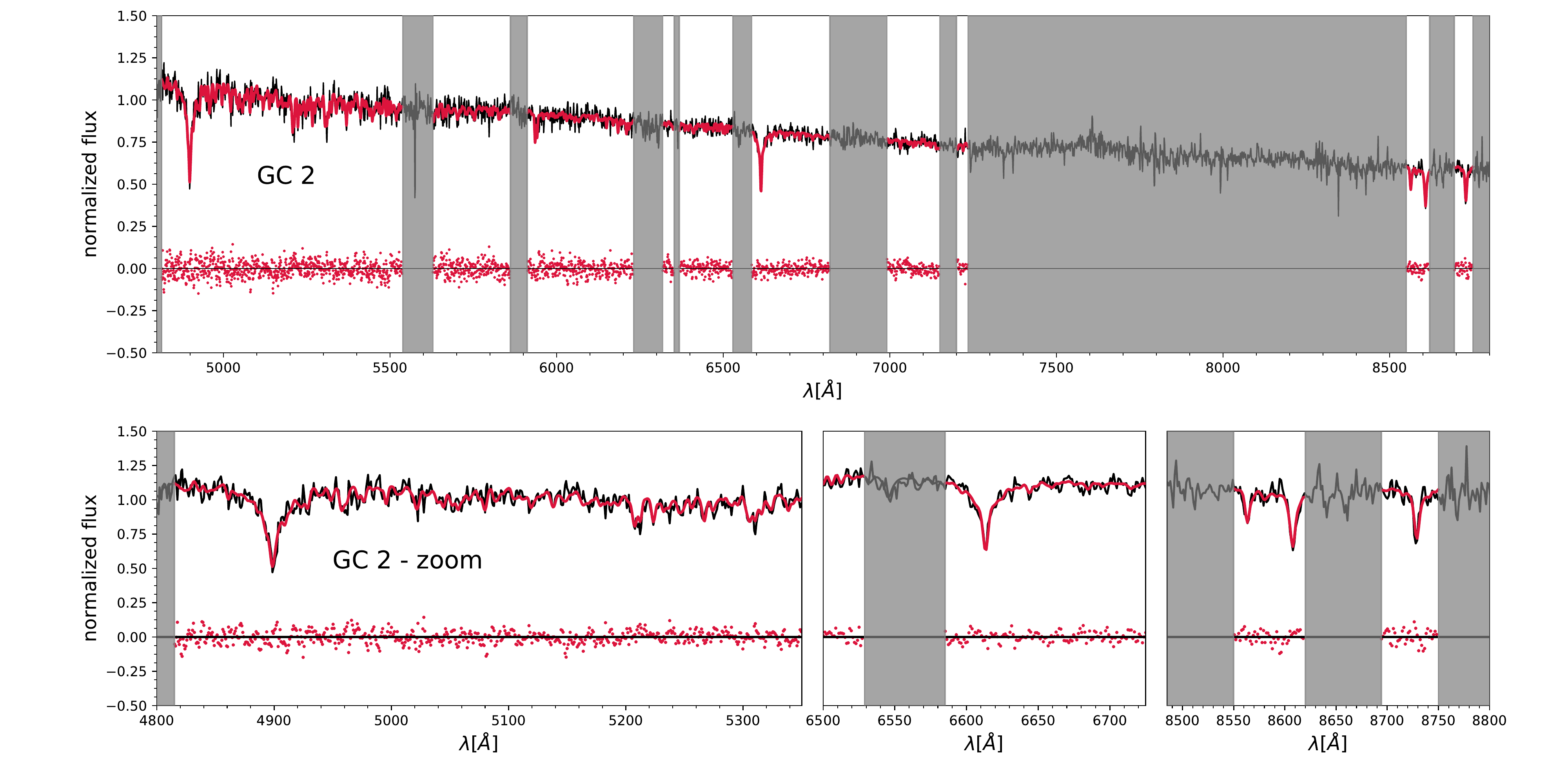}
    
      \caption{Comparison between the spectrum and the best SSP fit from pPXF for the shell and GC~2. The three plots in the bottom part of each panel show zooms on the important absorption lines. The gray regions are not taken into account for the fit. The scatter points show the residuals.
              }
         \label{fig:fits}
   \end{figure*}

   \begin{figure}[ht]
   \centering
   \includegraphics[angle=0,width=9cm]{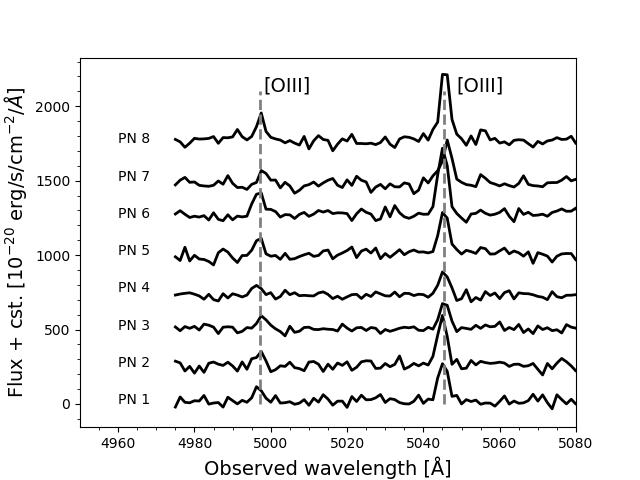}
      \caption{Spectra of the eight PNe candidates. The location of the [OIII] doublet at the redshift of NGC~474 is shown with gray dashed lines.
              }
         \label{fig:PNe}
   \end{figure}

We detected GCs and PNe using SExtractor on collapsed cubes \citep{Bertin96}. GC detections were done on the fully collapsed cube. We selected well defined point-like detections that had a velocity consistent with being part of this system, that is $\pm$ 400 km~s$^{-1}$ around the central velocity of NGC~474 and a S/N above 6~pix$^{-1}$, plus the candidates from L17. PNe were detected on stack of two collapsed cubes, each using only frames (slabs of $7.5~\AA$) around the redshifted wavelengths of each of the [O~\textsc{iii}] doublet emission line. To remove the continuum, we subtracted to this image two similarly collapsed cubes from nearby, but featureless, spectral regions. Eight sources with unambiguous detection of the [O~\textsc{iii}] doublet (S/N > 2.5) were detected, their location is indicated on Fig.~\ref{fig:map}. 

We extracted each point source with a Gaussian weight function to provide a S/N-optimised extraction. The full-width at half-maximum is chosen to be $\sim0.8\farcs$ to approximately match the resulting point spread function. The background is measured locally in eight nearby locations, using the same Gaussian weight. In each channel, we subtracted the median value of background regions to the source flux. The variance between those eight sky apertures is added to the source flux variance channel per channel. The flux from the shell is obtained by a sum of the flux on the full shell, after masking point sources, weighted by the white-light image to optimize the S/N.

The best single stellar population (SSP) fit for the shell and the brightest GC, GC~2, are shown in Fig.~\ref{fig:fits}. The gray regions are affected by residual sky lines and are not used for the fit. The other fits are presented in the Appendix and the methods are described in the next subsection. We note strong Balmer and calcium triplet (CaT) absorption lines but no emission lines. The S/N around the H$\alpha$ line is 65~pix$^{-1}$ for the shell, 26.9~pix$^{-1}$ for GC~2 and 6.6~pix$^{-1}$ for GC~3, respectively. The PNe candidate spectra are shown in Fig.~\ref{fig:PNe}. The S/N goes from 9.1 for PN~6 to 4.5 for PN~4.

\begin{table*}
\centering  
\caption{Results of the full spectra fitting procedure.\label{table1}. For the GCs, the S/N is measured in the neighborhood of the H$\alpha$ line and given in units of pix$^{-1}$. For the PNe, the S/N is integrated over the [OIII]$_{5007}$ line.}
\begin{tabular}{lccccccc}
    \hline\hline
  Name & RA (J2000)  & DEC (J2000) & S/N  & Velocity & Age  & M/H & GC: Mass [M$_\odot$]   \\
    &  [h:m:s] &  [h:m:s] &  &  [km.s$^{-1}$] &  [Gyr] &  & PNe: M$_{5007}$ [mag] \\
    \hline
    Shell & - & - & 65.0  & 2307.6 $\pm$ 2.7 & $3.55^{+0.61}_{-0.39}$  & $-0.83^{+0.12}_{-0.12}$   & - \\ \hline
    
    GC1   &  01:20:19.1987  &  03:25:49.4674  & 13.9   & 2294.7 $\pm$ 5.4 & $1.41^{+0.4}_{-0.22} $ & $ -1.07^{+0.17}_{-0.17}$  & $ 2.50 \times 10^4 \pm 5.89 \times 10^3$ \\
    GC2   &   01:20:19.2836  &  03:25:40.4404  & 26.9   & 2298.0 $\pm$ 4.8 &  $ 0.56^{+0.07}_{-0.06} $ & $ 0.22^{+0.22}_{-0.22}  $   & $ 5.91 \times 10^4 \pm 2.24 \times 10^3$ \\
    GC3   &    01:20:18.8817  &  03:25:34.9047 & 6.6    & 2307.6 $\pm$17.4 & -  & -  & - \\
    GC4   &  01:20:19.2141  &  03:25:20.0695  & 17.5   & 2304.0 $\pm$ 8.7 &  $ 12.59^{+2.17}_{-4.84} $ & $ -1.69^{+0.1}_{-0.18}  $ & $ 1.21 \times 10^5 \pm 2.21 \times 10^4$\\
    GC5   &   01:20:18.6095  &  03:24:40.3496   & 19.4   & 2267.5 $\pm$28.4 & $ 8.91^{+3.83}_{-2.78} $ & $ -1.73^{+0.13}_{-0.16}  $   & $ 1.35 \times 10^5 \pm 3.51 \times 10^4$\\
    GC6   &  01:20:19.1119  &  03:24:30.7437 & 9.4    & 2320.0 $\pm$49.5 & - &  - &  -\\ \hline

    PN1   &   01:20:18.7024  &  03:25:57.4425  & 4.1 & 2287.5 $\pm$ 6.6 & - & -  & $ -3.87 \pm 0.09  $ \\
    PN2   &  01:20:17.6471  &  03:25:28.4427  & 4.1 & 2275.6 $\pm$ 6.5 & - & -  & $ -4.10 \pm 0.08  $ \\
    PN3   &  01:20:19.7309  &  03:25:21.2427  & 4.8 & 2317.2 $\pm$ 6.3 & - & -  & $ -3.42 \pm 0.10  $ \\
    PN4   &  01:20:18.0078  &  03:25:13.2424  & 3.7 & 2294.5 $\pm$ 7.8 & - & -  & $ -3.32 \pm 0.11  $ \\
    PN5   &  01:20:18.3818  &  03:24:54.6427  & 4.2 & 2302.5 $\pm$ 6.8 & - & -  & $ -3.99 \pm 0.11  $ \\
    PN6   &  01:20:17.4201  &  03:24:48.0424  & 6.9 & 2286.0 $\pm$ 4.6 & - & -  & $ -4.46 \pm 0.07  $ \\
    PN7   &  01:20:18.8894  &  03:24:48.0424  & 2.8 & 2350.7 $\pm$ 8.3 & - & -  & $ -3.98 \pm 0.12  $ \\
    PN8   &  01:20:17.9410  &  03:24:47.6425   & 6.8 & 2316.8 $\pm$ 4.0 & - & -  & $ -4.51 \pm 0.08  $ \\
   \hline
   
\end{tabular}
\end{table*}


\section{Results}
\label{sec:res}

\subsection{Full spectra analysis}

The method is the same as the one described in \citet{Fensch19}. It is mainly based on pPXF \citep{Cappellari17} and the eMILES single stellar population (SSP) library \citep{Vazdekis16}. We linearly interpolate for sixteen more metallicity values, between [Fe/H]= -2.32 and -0.71, as done in \citet{Kuntschner10}. In the following we summarize the main points of the procedure. \\

For the shell and the GC candidates, we use pPXF with Legendre polynomials to account for uncertainty in the different flux calibration between our data and the eMILES library. For the kinematic fits, we use 12-degree multiplicative and 14-degree additive Legendre polynomials, and the full eMILES library, similar to \citet{Emsellem19}. For the stellar population fits, we use 12-degree multiplicative and no additive Legendre polynomials, and we individually fit with each SSP, as in \citet{Fensch19}. The ages we derive are then that of the best fitting single burst of star formation. Uncertainty on the results is obtained from a bootstrap method, consisting of resampling the residuals. This assumes that the residuals from the best fit are a good representation of the noise overall the spectrum, which is a reasonable assumption given the very small dependance of the residual spread with wavelength seen on Fig.~\ref{fig:fits}. We redistribute randomly the residuals from the best fit to obtain a residual spectrum that we add onto the best fit to create a re-noised spectrum. We repeat the operation 100 times to get a 100 re-noised spectra and the uncertainty is obtained as the dispersion in the new best fit value for the new 100 fits. For the PNe, we use a double Gaussian line emission template, with same width, to fit the [OIII] doublet. The uncertainty is obtained in the same way as for the shell and GCs. 
   
The result of the kinematic fits is shown in Fig.~\ref{fig:kin}. The GC and PN candidates have measured velocities within 50~km.s$^{-1}$ from the shell's velocity, as determined from the integrated MUSE spectrum. This kinematical association thus confirms a plausible association between them and the stellar shell. One should have in mind that the velocity profile of a stellar shell is expected to present a four-horn profile with a wide separation between the extreme values \citep[up to 150~km.s$^{-1}$, see e.g.][]{Ebrova12}. A spread of velocities of the PNe and GCs around the velocity of the integrated shell spectra is thus expected for such systems. We further note that the shell velocity, 2307.6~$\pm$ 2.7~km.s$^{-1}$ is similar to that of the central regions of NGC~474 \citep[$2315 \pm 5 $~km.s$^{-1}$][]{Cappellari11}. This fits within the interpretation that the shell is made of stars originating from the disrupted satellite that just reach the apocenter of their orbits in the potential well of the host galaxy: the sharp contrast of the shell emerges from a favorable projection, for which the apocenter, its motion, and the center of NGC~474 are on a plane aligned with the observer's sky plane \citep[see e.g.][]{Bilek16, Mancillas19simu}.\\

The results of the SSP analysis is shown in Fig.~\ref{fig:stel_pops} and are summarized in Table~\ref{table1}. The low S/N of GC~3 and GC~6, below 10~pix$^{-1}$, do not allow the fit from pPXF to converge in terms of stellar populations of these GCs and are not shown in that Figure. Our estimation gives the shell an intermediate age and metallicity, 
that is $3.55^{+0.61}_{-0.39}$~Gyr and [M/H] = $-0.83^{+0.12}_{-0.12}$. We note that these age and metallicity are significantly lower than that of the host galaxy as determined from the SAURON observations: $7.65 \pm 1.39$~Gyr and [M/H] = $-0.12 \pm 0.05$ within 1~R$_\mathrm{e}$ \citep{McDermid15}.

GC~4 and GC~5 are estimated to have a low metallicity and old age, above 9~Gyr, consistent with being old globular clusters. Their masses are estimated to be around $10^5$~M$_\odot$ from the fit. Their metallicity is similar to the so-called {\it blue} GCs \citep{Brodie06}. GC~1 and GC~2 are estimated to have a higher metallicity and much younger ages, around 1~Gyr. The estimated mass for these clusters are respectively $2.50 \times 10^4$~M$_\odot$ and $ 5.91 \times 10^4$~M$_\odot$.

   \begin{figure}[ht]
   \centering
   \includegraphics[angle=0,width=9cm]{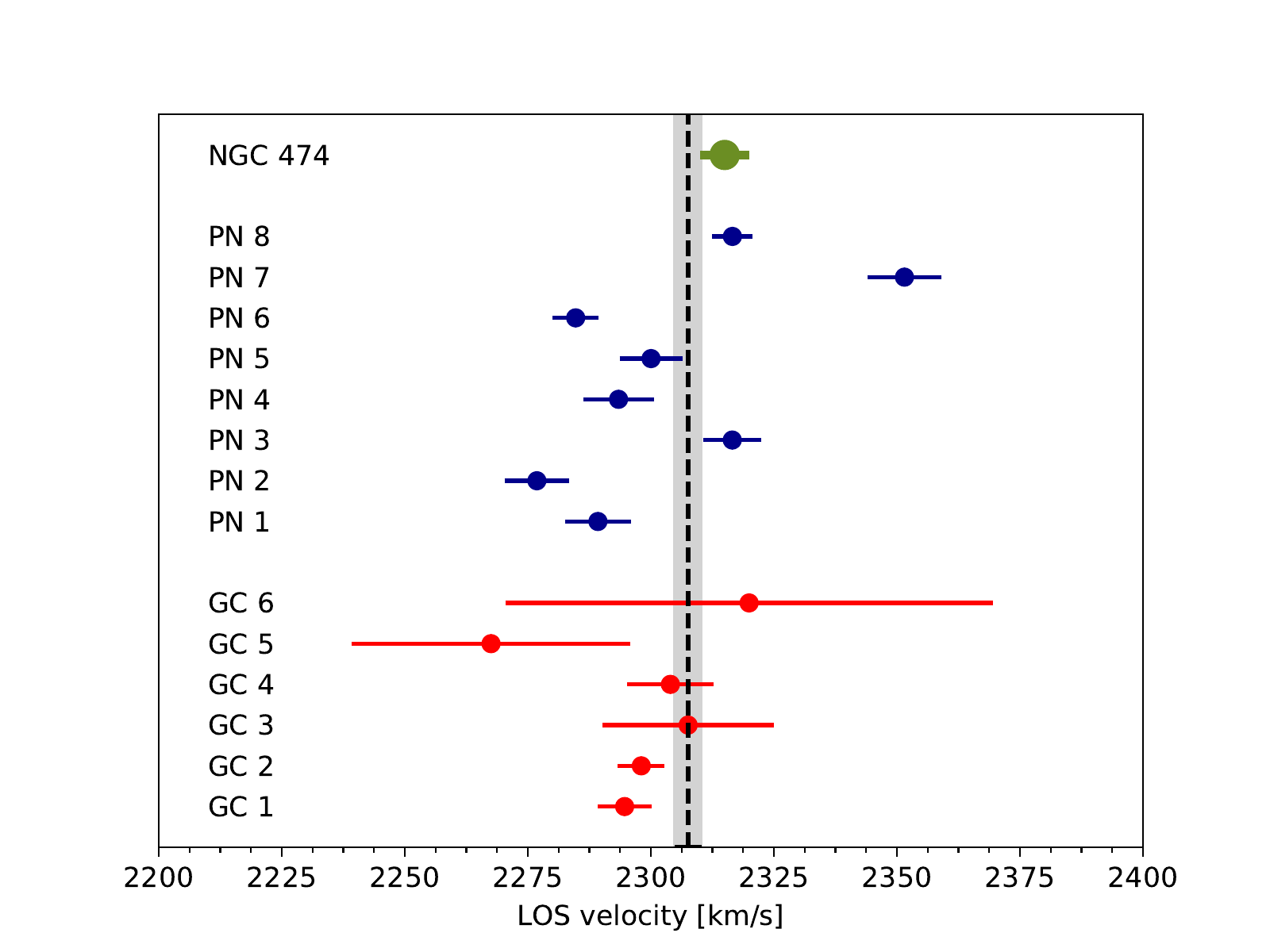}
      \caption{Line-of-sight velocity of the GCs, PNe and NGC~474. The black dashed line shows the line-of-sight velocity of the shell, and the gray shadow its uncertainty.
              }
         \label{fig:kin}
   \end{figure}

   \begin{figure}[ht]
   \centering
   \includegraphics[angle=0,width=9cm]{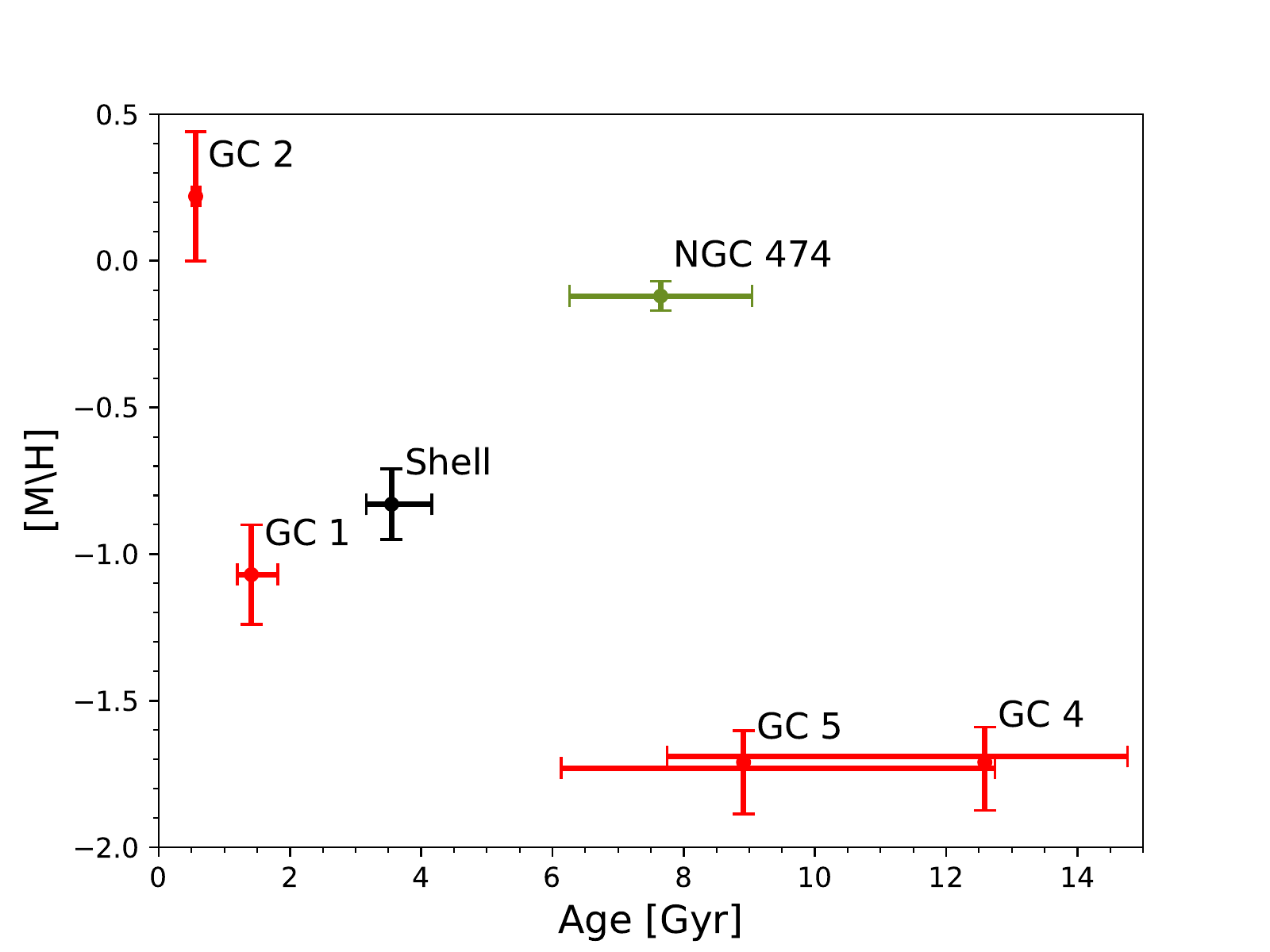}
      \caption{Estimated age and metallicity, assuming a SSP, for the shell and GC~1, GC~2, GC~4 and GC~5. The data point of GC~4 and GC~5 were slightly shifted for visualization purposes. The data point for NGC~474 (mass-weighted age within 1~Re) comes from \citet{McDermid15}.
              }
         \label{fig:stel_pops}
   \end{figure}

\subsection{Spectral indices}
\label{sec:lick}
As a complementary method, we estimate age, metallicities and $\alpha$ element enrichment from the standardized Lick/IDS system\footnote{We used the pyphot package available at this address: https://mfouesneau.github.io/docs/pyphot/index.html} \citep{Worthey94}. We show two diagnostic plots in Fig.~\ref{fig:Lick}, making use of the theoretical Lick indices for the MILES spectral library from \citet{Thomas10}. The error bars are the dispersion of the results of the Lick indices measurement on the re-noised spectra used for the measurement of the uncertainties in the previous section. We only show the result for the stellar shell, GC1 and GC2. The other objects have a too low SNR and thus have a significant fraction of their re-noised spectra with a negative equivalent width for at least one of the absorption lines of interest (more than 25\% against maximum one for the shell, GC1 and GC2). \\

The first diagnostic uses the Mg$b$ index and <Fe>, defined as the average between Fe5270 and Fe5335 \citep{Evstigneeva07}, which allow us to probe the $\alpha$-enrichment of the stellar population. For the ages considered (younger than 5~Gyr), we see that the spectral indices of these objects are consistent with a $\alpha$-enrichment [$\alpha$/Fe] = 0.3, but with rather large error bars due to the low SNR. We note that globular clusters are typically enriched to [$\alpha$/Fe] = 0.3-0.5 \citep[see review by][]{Brodie06}.

The second diagnostic uses the age-sensitive index H$\beta$ and the total metallicity-sensitive index [MgFe]' = $\sqrt{ \mathrm{Mg}b \times (0.72\mathrm{Fe}5270 + 0.28\mathrm{Fe}5335)}$ \citep{Evstigneeva07}. For GC~2, the two methods give very similar results. However, the Lick indices suggest a younger age and higher metallicities for the shell and GC~1 than our full fitting method. For the shell, Lick Indices suggest [M/H]$\sim$ -0.33 and an age between 1 and 3~Gyr. For GC~1, they suggest [M/H]$\sim$ -0.15 and an age around 0.8~Gyr but with large uncertainties. We note that Lick Indices do not show a metallicity difference between the two GCs, unlike the full spectral analysis. This second method confirms the trend that GC~1 and GC~2 are younger than the light-averaged stellar population in the stellar shell. We discuss the age difference between the two GCs in Section~\ref{disc:trigger}.

   \begin{figure}
   \centering
   \includegraphics[angle=0,width=9cm]{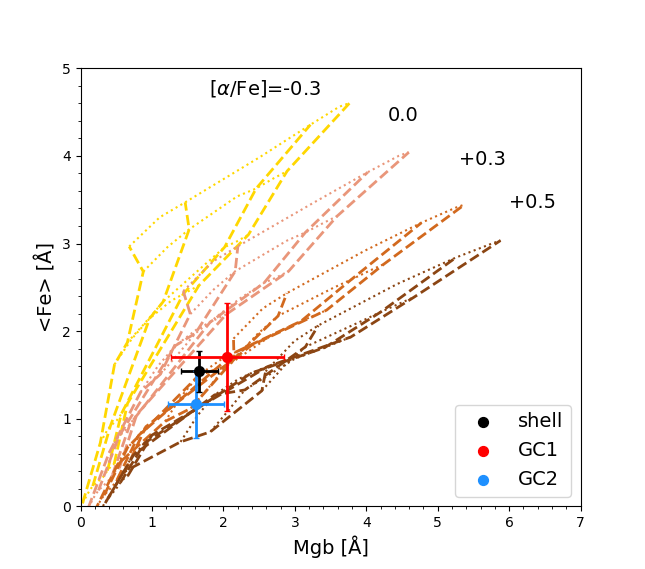}
   \includegraphics[angle=0,width=9cm]{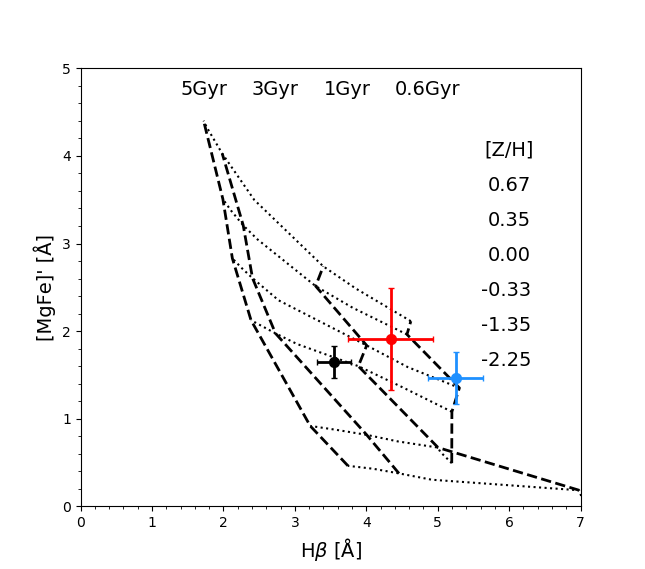}
      \caption{ Lick/IDS indices measured from the MUSE spectra of the shell, GC1 and GC2. Definition of the indices are given in the text. The model grids are from \citet{Thomas10} with isochrones of 0.6, 1, 3 and 5 Gyr in dashed lines and iso-metallicity of -2.25, -1.35, -0.33, 0.0, 0.35, and 0.67 dex in dotted lines. {\it Upper panel:} The grids have been color-coded with respect to $\alpha$-enrichment. The isochrones go from the leftmost (0.6~Gyr) to the rightmost one (5~Gyrs). {\it Lower panel:} The grid is shown for [$\alpha$/Fe] = 0.3. The isochrones go from the rightmost (0.6~Gyr) to the leftmost one (5~Gyr) and iso-metallicities go from the lowest one (-2.32) to the highest one (0.67).}
              
         \label{fig:Lick}
   \end{figure}

Last, the color-color diagram of detected GCs from L17's catalog is shown in Fig.~\ref{fig:ccd}. The estimated young ages of GC~1 and GC~2 is confirmed by their blue color (g-i < 0.5~mag), which excluded them from the L17 analysis. We note that GC~3 has similar (u-g, g-i) colors as GC~1 and GC~2, and was also rejected as GC candidates from the L17 analysis. However, its low S/N spectrum does not allow an estimate of its age and metallicity. These 3 GC candidates have bluer colors than the shell (g-i = 0.6~mag, L17). The origin of these GCs will be discussed in the following section.

   \begin{figure}[ht]
   \centering
   \includegraphics[angle=0,width=9cm]{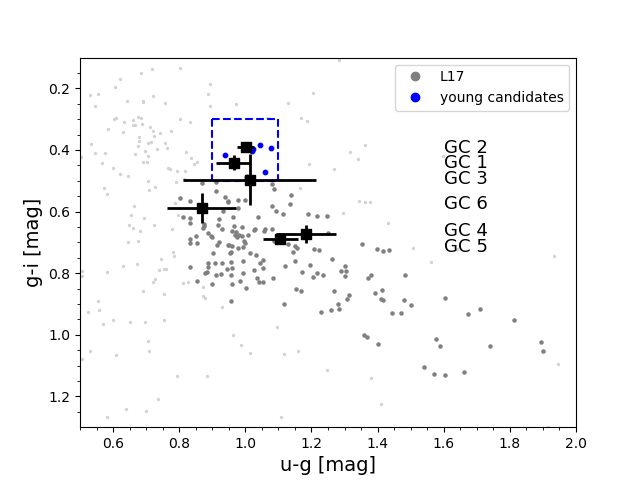}
      \caption{Color-color diagram of the point source detections in L17, corrected for foreground dust extinction. The thick points are the GC candidates selected by L17, while the light points were rejected. The 6 GCs within our field of view are shown as black squares, with their names on the right-hand side. The blue square and points show the selection of the GC candidates considered as young candidates in Section~\ref{sec:disc}.
         \label{fig:ccd}}
   \end{figure}

\section{Discussion}
\label{sec:disc}

\subsection{What is the progenitor of the shell?}

\begin{table}[]
\begin{tabular}{l | ccc}\hline\hline
 Constraint                                                                                           & Progenitor mass & Merger ratio \\ \hline
Stellar metallicity                                                                  &    $7.4^{+14.5}_{-4.7} \times 10^8$M$_\odot$           &      1: $109^{+195}_{-72}$        \\
\\
\begin{tabular}[c]{@{}l@{}}i-band luminosity and\\  15\% of stars in shell \end{tabular} &       $4.1^{+11.1}_{-2.9}\times10^{9}$~M$_\odot$           &     $ 1:21^{+51}_{-15} $      \\
\end{tabular}
\caption{ Progenitor mass and merger ratio estimates derived from a given constraint. See text for a description of the uncertainties considered in each case.}
\end{table}

Numerical simulations have shown that shell formation could emerge from relatively major mergers ( > 1:10 in stellar mass ratio, \citealt{Pop18, Karademir19}). One may wonder if the physical quantities we derived may constrain the mass of the progenitor of the shell. In the following we assume that the stars in the shell originate from a lower mass companion, the progenitor, and estimate the post-merger mass of NGC~474 to $10^{10.9}$~M$_\odot$ \citep{Cappellari13b}. \\

The estimated metallicity of the stars in the shell is  [M/H] = $-0.83^{+0.12}_{-0.12}$. Assuming that this value is a good estimation of the mean metallicity of the progenitor, i.e no strong initial metallicity gradients in the progenitor, the empirical stellar mass-metallicity relation gives an estimated mass of $7.4^{+14.5}_{-4.7}\times10^{8}$~M$_\odot$ \citep[derived from Equation 4 of ][accounting for the uncertainties from both their fit and our metallicity measurement]{Kirby13}. This would imply a merger ratio of 1:109 with $1\sigma$ lower and upper limits of respectively 1:305 and 1:37.

The stellar mass of a galaxy can also be roughly assessed independently via the number of PNe it hosts and its luminosity. The total number of PNe per bolometric luminosity in the brightest 1~mag of the luminosity distribution is typically $\alpha_1 = 3 \times 10^{-7} /40$ PNe/L$_\odot$ \citep[the factor 1/40 being derived for the canonical PN luminosity function]{Buzzoni06}. We have detected six PNe with M$_{5007}$ between -4.51 and -3.51 mag. Assuming that these six PNe are associated to the stellar population of the shell progenitor and not to the halo of NGC~474, which is reasonable given their similar line-of-sight velocity, we can estimate a bolometric luminosity of the shell of L$_{bol} \simeq 8 \times 10^8$~L$_\odot$. We note that the PNe systems around some shell galaxies outliers of typical scaling relations \citep{Buzzoni06}. \\

L17 have measured an absolute magnitude $M_g \sim -17.0$~mag and $M_i \sim -17.6$~mag for the shell. We assume a conservative error of 0.3~mag in both bands and use the color-dependant mass-to-light ratio from \citet{Roediger15}  with uncertainty 0.29 in solar units. Using $M_{\odot,i} = 4.53$~mag \citep{Willmer18}, one gets an estimation of the stellar mass in the shell of $5.6\times10^{8}$~M$_\odot$, with respective $1\sigma$ lower and upper limits: $1.7\times10^{8}$~M$_\odot$ and $1.9\times10^{9}$~M$_\odot$. We note that this estimation falls in the same ballpark as the rough estimate derived from the number of PNe.

Given that NGC~474 is surrounded by many shells and that the shells are expected to host only a fraction of the stellar light from the progenitor \citep[around 25 to 30\% from simulations, see][]{Hernquist92, Ebrova20}, the progenitor should be more massive than the above estimation which considered only the stellar shell. This also holds for the PNe estimation, assuming they follow the spatial distribution of the progenitor stars. Assuming that the shell studied here hosts $50 \pm 10 \%$ of the stars in the outer shell system created by the interaction, (See Fig.~\ref{fig:map}), and assuming that the shells host $30 \pm 10\%$ of the progenitor stellar mass, the luminosity gives an estimated progenitor mass of $4.1^{+11.1}_{-2.9}\times10^{9}$~M$_\odot$. This translates into a merger ratio of 1:21, with respective lower and upper $\sigma$ limits of 1:72 and 1:6.

The estimation from the stellar metallicity and the luminosity differ by a factor five, but given the large error bars, the two values are inconsistent by only $\sim 1~\sigma$. Nevertheless, one should note that we have so far assumed that the progenitor has a constant metallicity, which is a disputable hypothesis. \\

Resolved metallicity gradients in local star-forming galaxies from the CALIFA survey show that a stellar metallicity as low as [M/H] = - 0.83 is attained out to 2.5 effective radius in galaxies with stellar mass above 10$^{9.1}$~M$_\odot$ only for rare outliers, the constrain being stricter in groups \citep{Gonzalez-Delgado15, Coenda20}. However, one should note that other surveys, such as MaNGA, did find steeper metallicity gradients in their local star-forming galaxy sample, with metallicity as low as [M/H] = -0.7 at 1.5 effective radius even for 10$^{10.5-11}$~M$_\odot$ galaxies \citep{Lian18}. \\

Thus, on the one hand, the observed stellar shell may have formed via a $\sim 1:100$ merger event, and in this case the observed shell would contain most of the stars of the progenitor, contrarily to the 15\% assumed here. The various outer shells would then come from different progenitors, including possibly the primary galaxy. 

On the other hand, the observed stellar shell may have formed via a $\sim 1:20$ merger event, from a galaxy with a few $10^9$~M$_\odot$ stellar mass and lower mean metallicity than typical galaxies of similar stellar mass, or with a metallicity gradient. \\

 These two scenarios might be further explored using spectroscopic data of the other shells and a chemo-dynamical numerical model of the collision. We further note that \citet{Alabi20}\footnote{This article was made public during the refereeing process of the present article.} suggest a more massive progenitor, namely $\simeq 10^{10}$~M$_\odot$, based on a higher metallicity measurement. One would thus need observations revealing the kinematics and the stellar populations -- age and metallicity) -- within the other shells, for instance using integral field spectroscopy. Given the low surface brightness of the shells and the long exposure time on a 8-meter class telescope used for this study, we estimate that such a study will only be made possible by the upcoming generation of 30-meter class optical telescopes. Once these measures are obtained, numerical simulations using different progenitor masses and metallicity gradients would be needed to show which of the proposed configurations are actually achievable within the observations constraints.

\subsection{Could the shell forming event trigger massive star cluster formation?}
\label{disc:trigger}

From cosmological zoom simulations, \citet{Mancillas19simu} have estimated that stellar shells can remain visible for 3 to 4~Gyr. In Section~\ref{sec:res}, we estimated that GC~1 and GC~2 have ages younger than 1.5~Gyr. One may thus wonder if they were formed during the event that formed the stellar shell. 

Observations and simulations of interactions of galaxies have shown that such events could trigger the formation of star clusters, similar in mass to GC~1 and GC~2 \citep[see e.g.][]{Whitmore10, Renaud15, Lahen19}. \citet{Sikkema06} have found hints for newly formed GCs in the nucleus of two shell galaxies (type II NGC~2865 and type III NGC~7626), but not for NGC~474. We have noted in Section~\ref{sec:res} that GC~1 and GC~2 have different ages and metallicity. To explain that, one may estimate the free-fall time at the radius of the shell, that is 30~kpc. Via abundance matching \citep{Behroozi13, Diemer15} one may estimate that the most likely halo to host NGC~474 would have a total mass of $9 \times 10^{11}$~M$_\odot$ and a virial radius of 1200~kpc, which gives a free-fall time of about 100~Myr. This means that the GCs have already passed close to the nucleus of NGC~474 several time since their formation, and makes it possible that their formation happened at different passages of the associated material close to that nucleus.\\

One way to test the hypothesis of triggered star cluster formation is to check whether we see traces of star formation from the same epoch in the nucleus of NGC~474, where we may expect the build-up of a gas reservoir leading to a potential starburst.

We note that \citet{Jeong12} and \citet{Zaritsky14} classified NGC~474 as a {\it blue outlier} among early-type and lenticular galaxies because of its relatively blue color and elevated near-UV emission. Moreover, \citet{Sikkema07} noticed pronounced dust lanes (estimated mass of $\sim$ 10$^4$~M$_\odot$) in the central 15$\arcsec$ of NGC~474, which align with the location of ionized gas. No molecular gas was found in the central region of NGC~474 \citep[upper limit: M(H$_2$) < $10^{7.7}$~M$_\odot$, see ][]{Combes07, Mancillas19gas}. \citet{Schiminovich97} and \citet{Rampazzo06} found an HI tail due to the interaction with NGC~470 (spiral galaxy westward of NGC~474 in Fig.~\ref{fig:map}), but it is not superimposed with the shell.\\

To investigate the presence of a young stellar population in the host galaxy, we use the central kpc of the SAURON data cube centered on NGC~474 from \citet{Cappellari11} and used the pPXF regularization procedure from \citet{Cappellari17}, with 3-degree multiplicative Legendre polynomials and the previous eMILES library. To avoid degeneracies due to the presence of emission lines, namely H$_\beta$ and the [OIII] doublet, we perform a first fit by masking the location of these lines and fix the obtained velocity and velocity dispersion for a second fit enabling emission lines. Uncertainties on the fit are obtained with the bootstrap method used for the MUSE data.

The regularized fit to the data and the weights of the different templates are shown respectively in the upper and lower panel of Fig.~\ref{fig:sauron_fit}. We stress that the full SSP library is considered by pPXF, but only some of them get a non-zero weight from the regularization procedure. On top of old and metal-rich templates, typical of elliptical galaxies, the regularization gives a non-zero weight to templates of age and metallicity similar to GC~1 and GC~2 (see Fig.~\ref{fig:stel_pops}). The regularization also gave non-zero weights to very metal-poor templates, at the limit of our metallicity grid. To estimate the significance of these metal-poor templates on our estimation, we create two templates: {\it Old, high Z} and {\it Old + Young, high Z} which respectively contains the old and metal-rich (age > 7~Gyr and [M/H] > -2), and all the metal-rich ([M/H] > -2) templates used in the regularized fit with their respective weights. Then we fit the SAURON spectrum with each of these templates, following the same procedure as above. The results are shown in Fig.~\ref{fig:sauron_fit_test}. We see that the {\it Old, high Z} population alone struggles particularly to account for the region around the H$_\beta$ and Mgb absorption lines. Adding the young and metal-rich population improves significantly the fit, especially around the H$_\beta$ line. Adding this population reduces the $\chi^2$ by a factor 2.4. Adding the metal-poor population improves the fit around the Fe lines (observed wavelength $\sim 5054$ $\AA$ for l.o.s velocity of 2315 km~s$^{-1}$) while overestimating the flux in the Mgb line. Adding this population reduces the $\chi^2$ by a 10\%. 

   \begin{figure}[ht]
   \centering
   \includegraphics[angle=0,width=9cm]{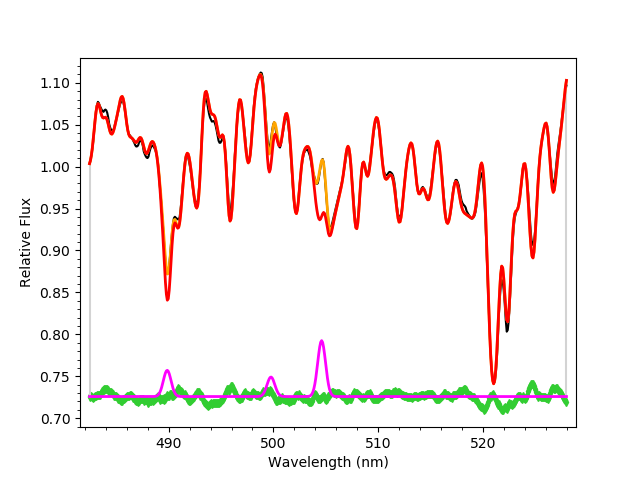}
    \includegraphics[angle=0,width=9cm]{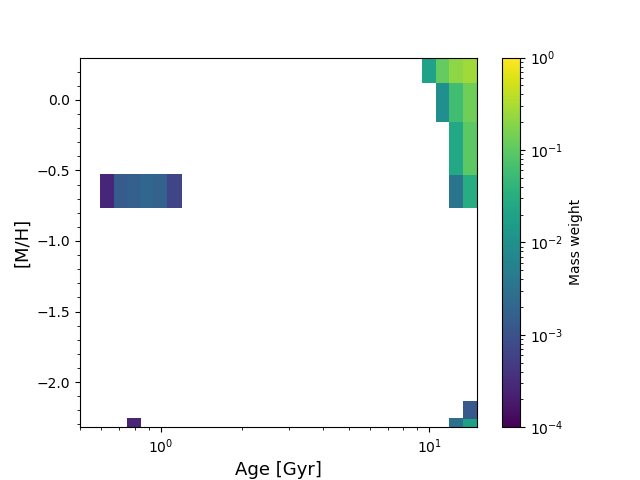}
      \caption{Upper panel: Best fit for the SAURON data of NGC~474 (central kpc), from the pPXF procedure. The data is in black, the fit in red for the star component and yellow for the gas component. The green points show the residuals and the pink line highlight the ionized gas component fit: from left to right H$\beta$ and the [OIII] doublet.
      Lower panel: weights of the different templates used by the regularized fit. Only colored regions are contained in the resulting fit. See text for details.
              }
         \label{fig:sauron_fit}
   \end{figure}
   
    \begin{figure}[ht]
   \centering
   \includegraphics[angle=0,width=9cm]{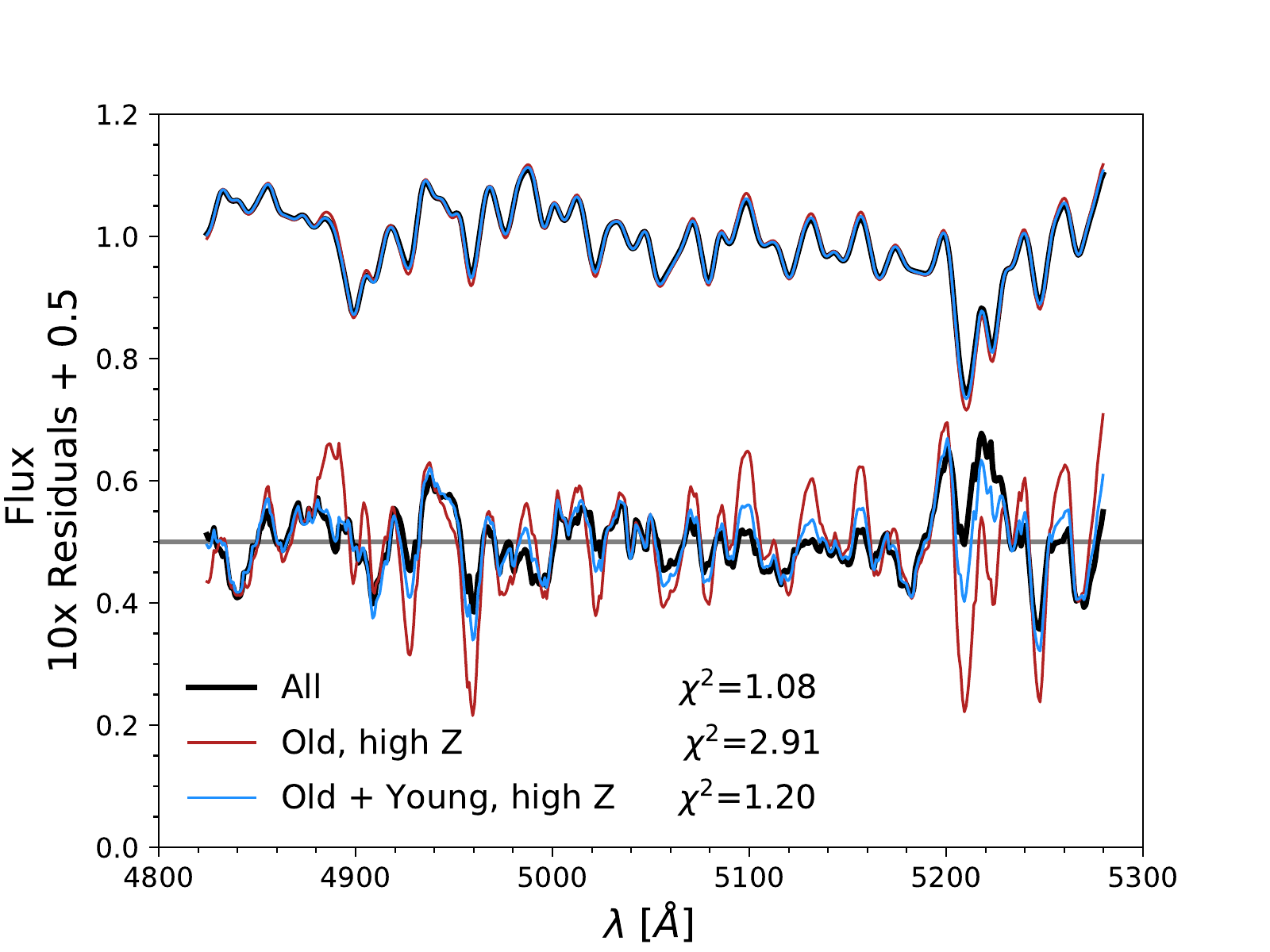}
      \caption{Best fits, residuals and $\chi^2$ values corresponding to the procedure described in the text. The residuals have been multiplied by 10 and shifted by a constant 0.5 offset for the sake of visualization. Flux unit is arbitrary.
              }
         \label{fig:sauron_fit_test}
   \end{figure}

The fit infers a total mass for the young and metal-rich component of $3.22 \pm 1.37 \times 10^{7}$~M$_\odot$, to be compared with $4.00 \pm 0.83 \times 10^{9}$~M$_\odot$ for the old and metal-rich component. The young (resp. old) and metal-poor component in the fit is evaluated to 3\% (resp. 2.4 \%) of the young (resp. old) and metal-rich in terms of mass: $1.07 \times 10^6$ M$_\odot$ (resp. $9.52 \times 10^7$ M$_\odot$). 

Given the low level of improvement of the fit by adding this very metal-poor population and their low relative mass contribution, we discard them in the following. The inclusion of these populations at the limit of our metallicity grid and for the same ages as the most important contributors to the fit might be due to the imperfection of the used SSP models, in particular in terms of [$\alpha$ / Fe] \citep[we only use 'base models', see discussion in ][]{Vazdekis16}.

The SAURON spectrum of NGC~474 therefore supports the hypothesis of a nuclear starburst around 1~Gyr ago. We note that this timescale is similar to the estimated ages of the massive stellar clusters, and suggests that both the starburst and cluster formation could have been caused by a single galaxy merger. However, we warn that the structure of NGC~474 is complex \citep[see e.g.][]{Sikkema07} and that this young stellar component cannot yet be unambiguously connected to the shell. 

One may thus wonder what is the origin of the gas that fueled this starburst and the formation of the GCs. Indeed, the nucleus of NGC~474 does not host any detectable H$_2$ and we note that no HI has been detected at the vicinity of the shell \citep{Schiminovich97, Rampazzo06, Mancillas19gas}. This suggests that these episodes of star formation were very efficient at consuming or ejecting the gas and calls for a dedicated numerical study of star cluster formation in shell-forming galaxy interactions.  \\

Last, we locate the GC candidates all around NGC~474 (see field of view in Fig.~\ref{fig:blue}) with similar colors as confirmed young GC~1 and 2. We chose GC candidates with 0.3 < g-i < 0.5 and 0.9 < u-g < 1.1, depicted as {\it young candidates} in Fig.~\ref{fig:ccd}. There are nine young candidates in total, including GC~1, 2 and 3. There is one more young candidate in the MUSE field-of-view. After verification, this source was discarded from the present analysis because of its very low S/N = 2.8 pix$^{-1}$, which did not permit any reliable analysis. Apart from GC~1 and GC~2 (with resp. $m_g = 23.04 \pm 0.02$~mag and $22.19 \pm 0.01$~mag), these candidates have $m_g$ between 24 and 25~mag. We note that their relative faintness would not allow age and metallicity measurement using MUSE with less than 10.2~h of on-source integration. These six young candidates are shown with cyan circles in Fig.~\ref{fig:blue}. We note that they are located in the vicinity of the studied shell, inner stellar shells or tidal features, suggesting a possible link with the events that created these structures.

   \begin{figure}[ht]
   \centering
   \includegraphics[angle=0,width=9cm]{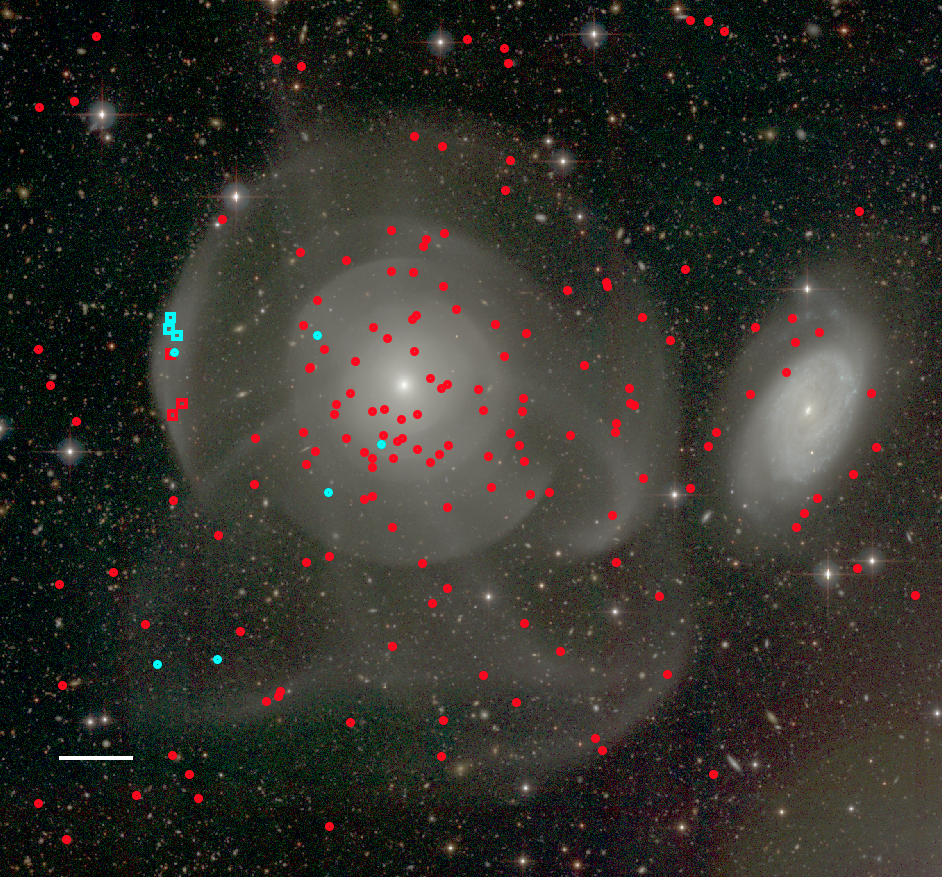}
      \caption{True color image from Megacam@CFHT with location of the GC candidates around NGC~474. Young candidates are shown in cyan, old candidates in red. The GC candidates for which we have MUSE spectroscopy are shown with squares. The white line spans 1 arcmin.
         \label{fig:blue}}
   \end{figure}

 \section{Conclusion}
 \label{sec:conc}
 
We use MUSE@VLT to study the origin of an external stellar shell around the host galaxy NGC~474. 

We find eight PNe and six GC candidates (including three candidates from \citealt{Lim17}) which are kinematically associated with the stellar shell. Stars from the shell have a luminosity-weighted age younger than that of the host, $3.55^{+0.61}_{-0.39}$~Gyr. We show that the shell has an intermediate metallicity, [M/H] = $-0.83^{+0.12}_{-0.12}$ and a possible $\alpha$-enrichment, [$\alpha$/Fe] $\sim$ 0.3.\\

Assuming the material of the shell comes from a lower mass companion, and that the latter had no initial metallicity gradient, the measured stellar metallicity would constrain the mass of the progenitor to be around 7.4$\times 10^8$~M$_\odot$, implying a merger mass ratio of about 1:100. However our census of PNs and archival photometry of the shell would suggest a higher merger ratio, namely 1:20. While this discrepancy is significant only at the $1~\sigma$ level, different scenarios are discussed, namely that the shell contains a high proportion of the stars from the progenitor, that the progenitor has a lower stellar metallicity than typical for a given stellar mass or that the stars of the shell originate from low metallicity outskirts of a more massive galaxy with a metallicity gradient. Each of these scenarios could be assessed with dedicated chemo-dynamical observations and numerical simulations. \\

Full spectral fitting of the four GC candidates with high enough S/N shows that two of them are consistent with being old globular clusters while the other two show signs of young ages (<1.5~Gyrs) and relatively high metallicity ([M/H] > -1.1). The study of the Lick Indices confirms the trend that GC~1 and GC~2 are younger than the stars in the shell. Their different age and metallicity hint to a formation at different passage of the associated material close to the nucleus. This is consistent with their blue color, which excluded them from previous analysis of GCs in this system. Regularization of full spectral fitting on SAURON data of the central kpc of NGC~474 hints for the presence of a relatively young stellar population, with age and metallicity similar to that of these young GC candidates. This suggests that these young GC candidates might have formed during the merging event that causes the formation of the stellar shell. We note the presence of nine GC candidates with similar color around NGC~474, all fainter that GC~2, the brightest young cluster candidate, which is estimated to have a mass of $\sim 6 \times 10^4$~M$_\odot$. \\

This data thus supports the scenario in which these stellar shells have formed from a progenitor with overall low-metallicity or from its low metallicity outskirts, and during a merger that might have possibly triggered the formation of massive star clusters and/or triggered a nuclear starburst in the host. This calls for dedicated numerical studies to confirm or disprove such a scenario.

\begin{acknowledgements}
We thank the referee for insightful comments that help improve this paper. JF would like to thank the ESO User Support Department for their great help on performing these observations. The authors thank Richard McDermid and Harald Kuntschner for interesting discussions respectively about NGC~474 and Lick Indices. C.L. acknowledges support from the National Natural Science Foundation of China (NSFC, Grant No. 11673017, 11833005, 11933003, 11621303, 11973033 and 11203017)
\end{acknowledgements}
\bibliographystyle{aa}  
\bibliography{library}

\begin{appendix}
\section{GC spectra and fits}

\begin{figure*}
   \centering
   \includegraphics[angle=0,width=18cm]{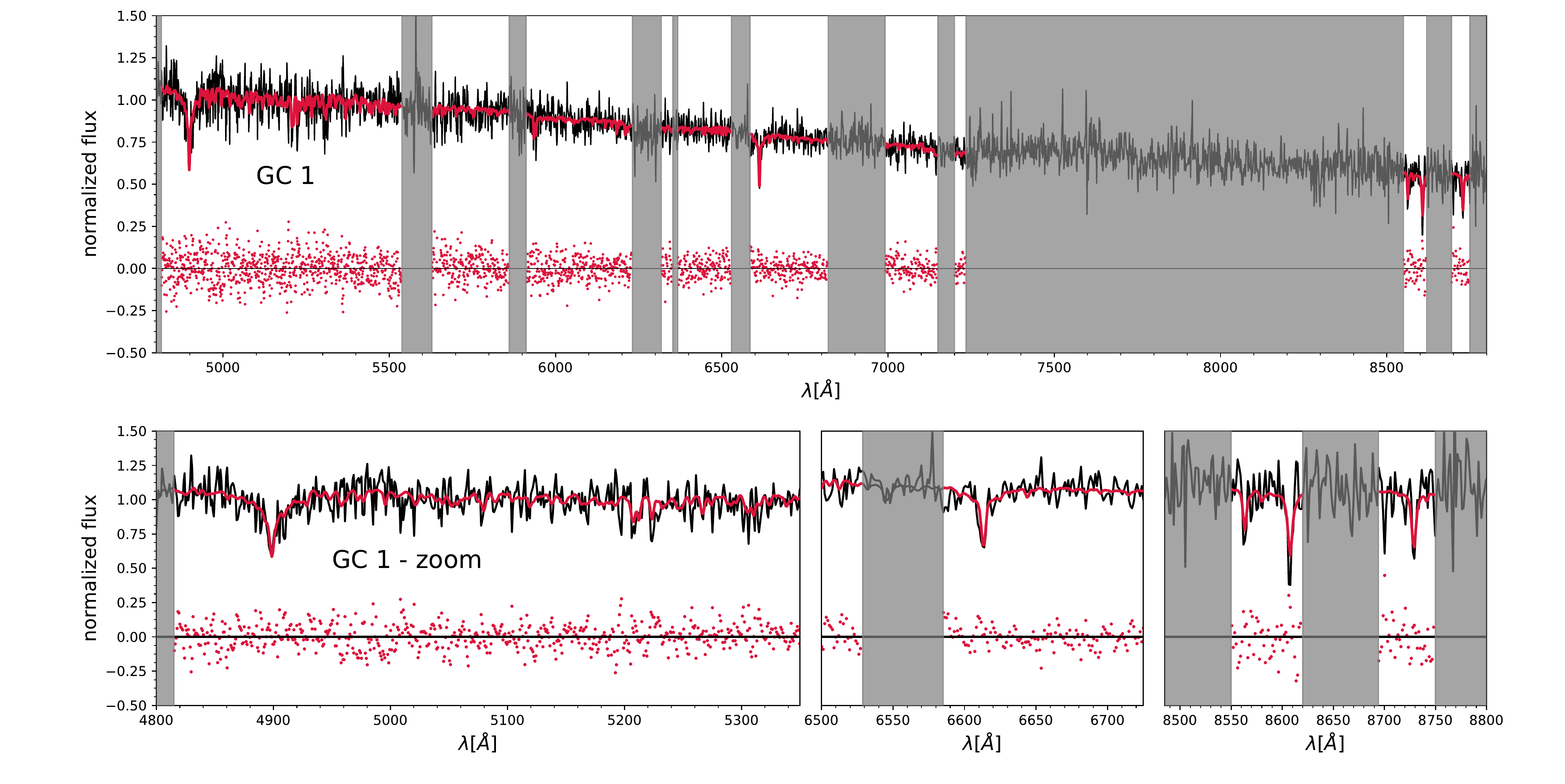}
   \caption{Same as Fig.2}
\end{figure*}

\begin{figure*}
   \centering
   \includegraphics[angle=0,width=18cm]{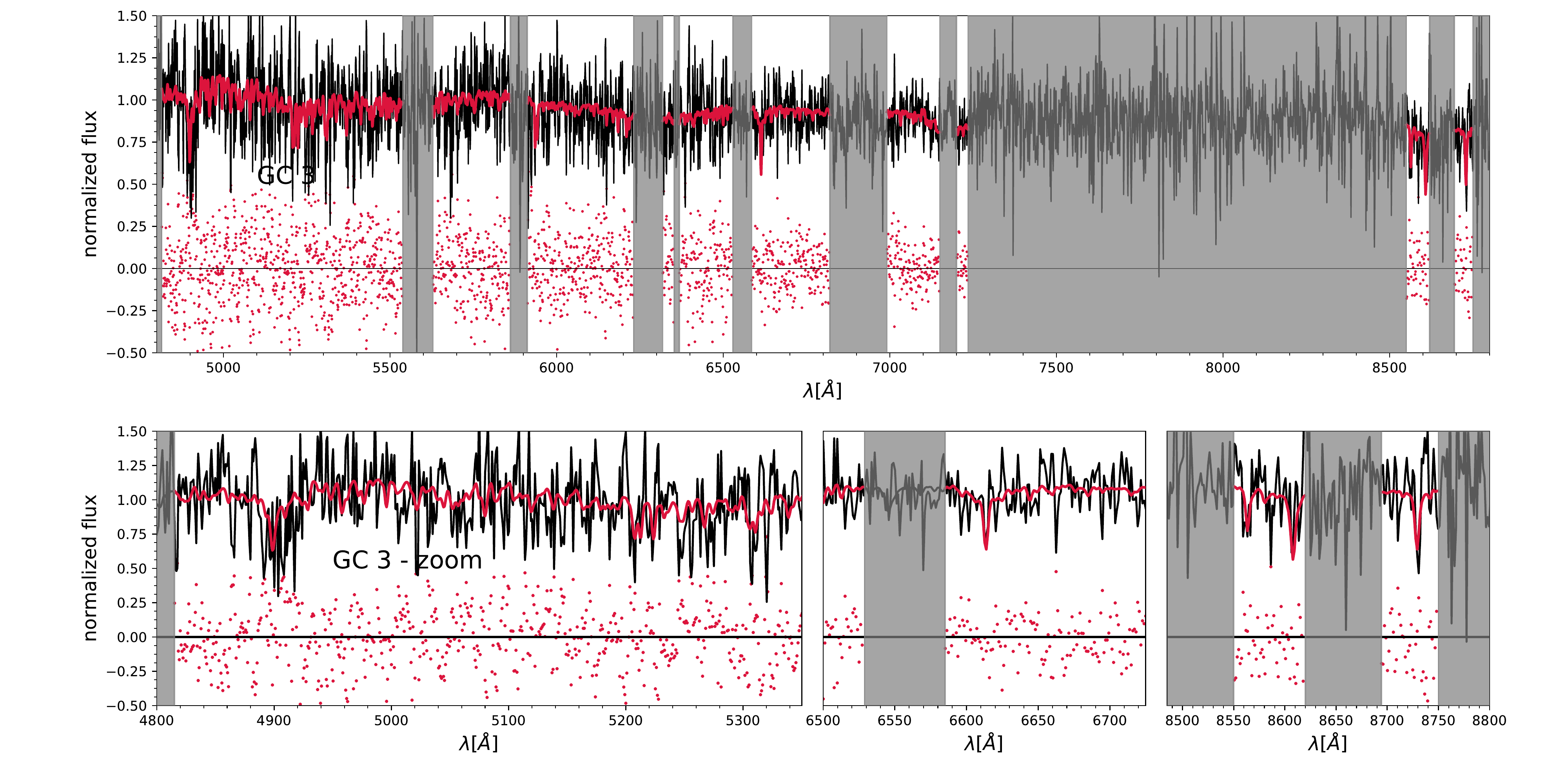}
      \caption{Same as Fig.2}
\end{figure*}
   
\begin{figure*}
   \centering
   \includegraphics[angle=0,width=18cm]{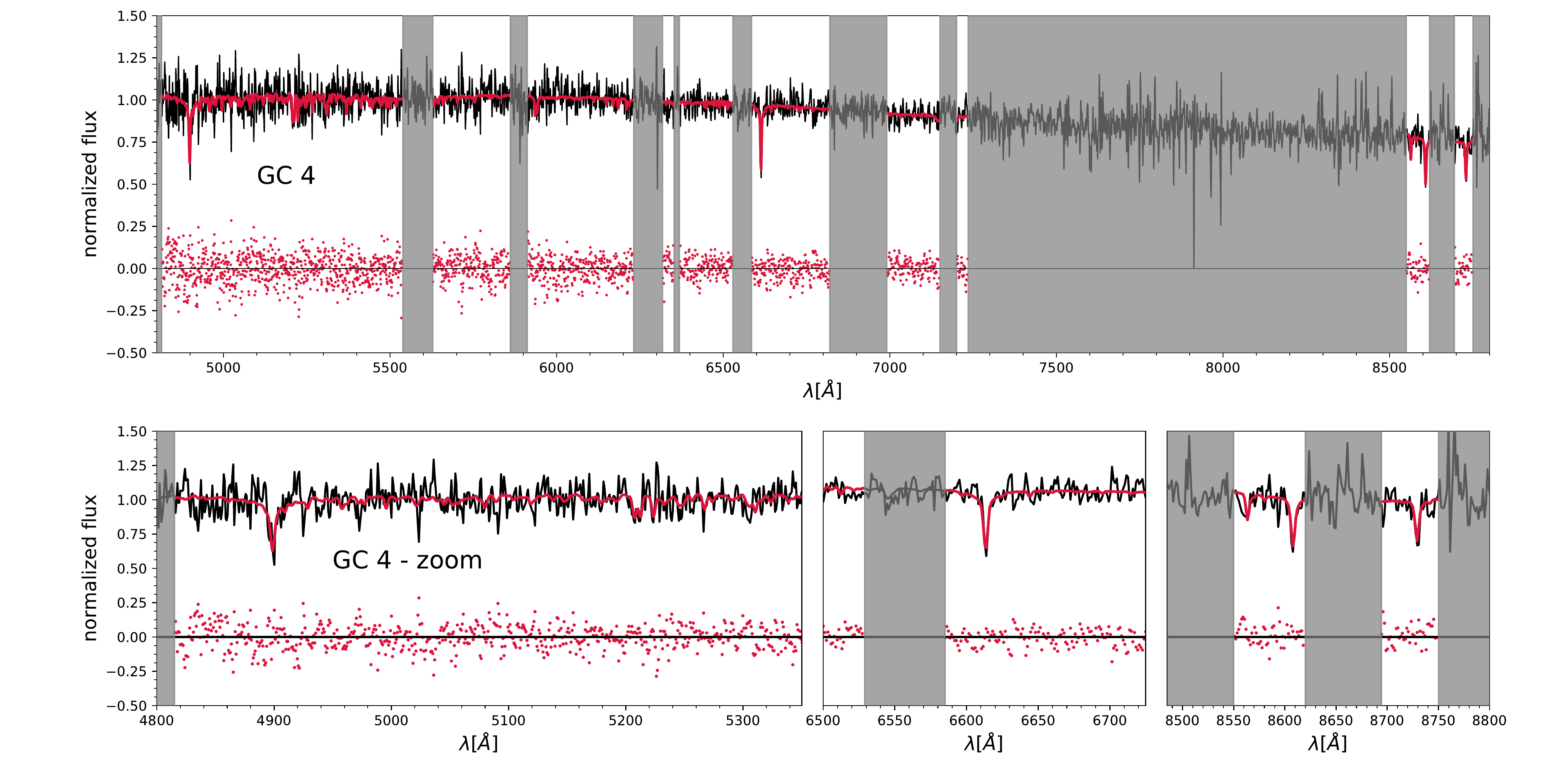}
      \caption{Same as Fig.2}
\end{figure*}

\begin{figure*}[ht]
   \centering
   \includegraphics[angle=0,width=18cm]{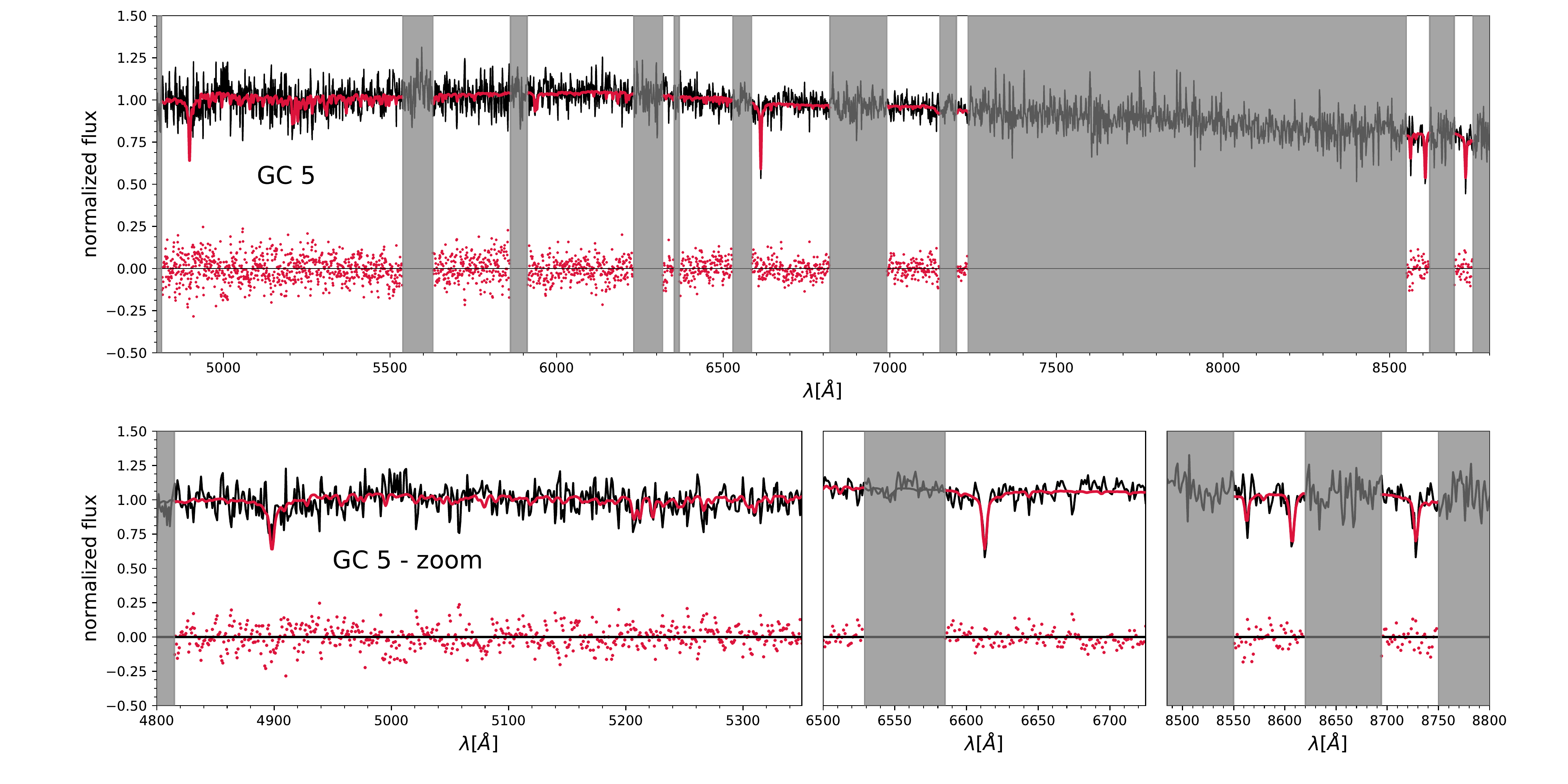}
      \caption{Same as Fig.2}
\end{figure*}

\begin{figure*}[ht]
   \centering
   \includegraphics[angle=0,width=18cm]{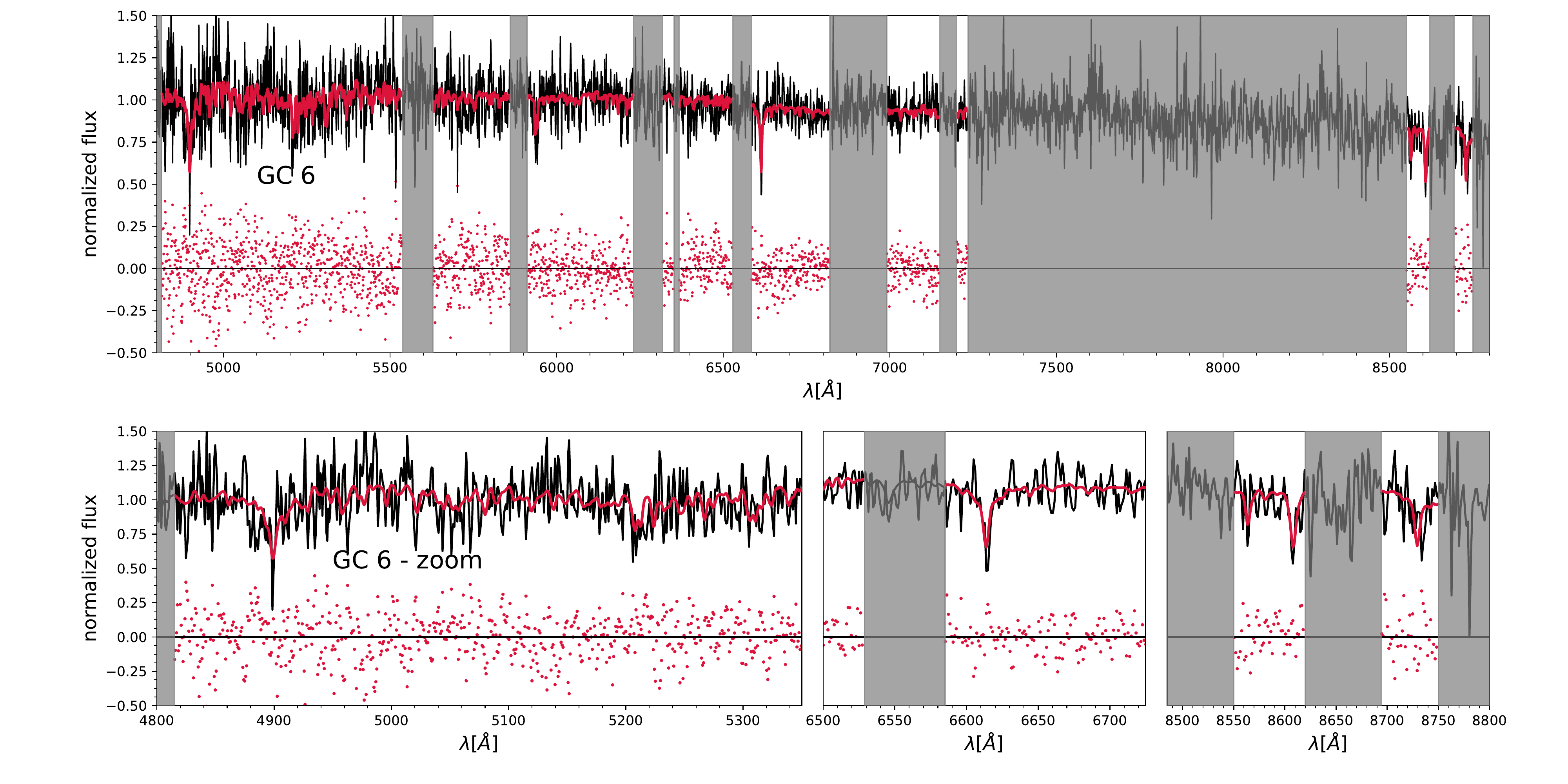}
      \caption{Same as Fig.2}
\end{figure*}

\end{appendix}

\end{document}